\documentclass[journal, twoside]{IEEEtran}

\ifCLASSINFOpdf
    \usepackage[pdftex]{graphicx}
\else
\fi
\usepackage[cmex10]{amsmath}
\usepackage{pgfplots}
\usetikzlibrary{calc}

\begin{document}
%\pagestyle{empty}
\title{A $z$-Vertex Trigger for Belle~II}

\author{
S. Skambraks, 
F. Abudin\'en, Y. Chen,  M. Feindt, R. Fr\"uhwirth, M. Heck, C.~Kiesling, A. Knoll, S. Neuhaus, S.~Paul, J. Schieck

\thanks{Manuscript received June 13, 2014; revised April 30, 2015; accepted May~8, 2015.}
\thanks{This work was supported by the DFG Cluster of Excellence
``Origin and Structure of the Universe''. }
\thanks{Stefan Paul and Sebastian Skambraks as corresponding author are with the Physik Department, Technische Universit\"at M\"unchen, Germany (e-mail: sebastian.skambraks@tum.de)}
\thanks{ Yang Chen, Alois Knoll and Sara Neuhaus are with the Institute for Robotics and Embedded Systems, Technische Universit\"at M\"unchen, Germany} 
\thanks{ Fernando Abudinen and Christian Kiesling are with the Max-Planck-Institut f\"ur Physik,
Germany} 
\thanks{Martin Heck and Michael Feindt are with the Institute for Experimental Nuclear Physics, Karlsruher Institut f\"ur Technologie, Germany} 
\thanks{Rudolf Fr\"uhwirth and Jochen Schieck are with the Institute for High Energy Physics, \"Osterreichische Akademie der Wissenschaften, Austria}

%		Michael~Shell,~\IEEEmembership{Member,~IEEE,}
 %       John~Doe,~\IEEEmembership{Fellow,~OSA,}
  %      and~Jane~Doe,~\IEEEmembership{Life~Fellow,~IEEE}% <-this % stops a space
%\thanks{C2PAP computing cluster - excellence cluster universe}
}
\IEEEpubid{\copyright~2014 IEEE}


% invoke bibstyle changes
\bstctlcite{IEEEexample:BSTcontrol}

\maketitle

\begin{abstract}
%\boldmath
The Belle~II experiment will go into operation at the upgraded
SuperKEKB collider in 2016. SuperKEKB is designed to deliver an
instantaneous  luminosity
$\mathcal{L}=8\times10^{35}\,\mathrm{cm}^{-2}\,\mathrm{s}^{-1}$.
The experiment will therefore have to cope with a much larger
machine background than its predecessor Belle, in particular from
events outside of the interaction region. We present the concept
of a track trigger, based on a neural network approach, that is
able to suppress a large fraction of this background by
reconstructing the $z$ (longitudinal) position of the event vertex
within the latency of the first level trigger.
 
The trigger uses the hit information from the Central Drift
Chamber (CDC) of Belle~II within narrow cones in polar and
azimuthal angle as well as in transverse momentum (``sectors''),
and estimates the $z$-vertex without explicit track
reconstruction. The preprocessing for the track trigger is based
on the track information provided by the standard CDC trigger. It
takes input from the 2D track finder, adds information from the
stereo wires of the CDC, and finds the appropriate sectors in the
CDC for each track.

Within the sector, the $z$-vertex is estimated by a specialized
neural network, with the drift times from the CDC as input and a
continuous output corresponding to the scaled $z$-vertex.

The neural algorithm will be implemented in programmable hardware. 
To this end a Virtex~7 FPGA board will be used, which provides at
present the most promising solution for a fully parallelized
implementation of neural networks or alternative multivariate
methods. A high speed interface for external memory will be
integrated into the platform, to be able to store the
$\mathcal{O}(10^9)$ parameters required.

The contribution presents the results of our feasibility studies
and discusses the details of the envisaged hardware solution.

\end{abstract}

\begin{IEEEkeywords}
%IEEEtran, journal, \LaTeX, paper, template.
Trigger, Neural Networks, CDC, Belle~II, SuperKEKB, MLP, L1
\end{IEEEkeywords}

% \fi
\IEEEpeerreviewmaketitle

\section{Introduction}
\IEEEPARstart{N}{eural} Networks of the Multi Layer Perceptron
(MLP) type, implemented as a $z$-vertex predictor, are the key
concept of our proposed first level (L1) track trigger system for the
Belle~II experiment~\cite{SSDA, SSMA}.

This track trigger system is designed to suppress background
events with vertices outside the interaction region. 
Using only the hits in the Central Drift Chamber (CDC), an
estimation of the $z$-position of the track vertex is made by a
neural net approach without explicit track reconstruction. Such a
machine learning approach is superior to an analytic solution in
the aspect of noise robustness~\cite{FAMA} and in its
deterministic runtime.

The trigger has to respect the requirements of the
L1 trigger
of Belle~II, especially the total latency of
$5\,\mu\mathrm{s}$ for the full L1 trigger system and the required
final trigger rate of $30\,\mathrm{kHz}$ at a minimum two event
separation of $200\,\mathrm{ns}$~\cite{iwasaki}.
We propose an implementation in FPGAs which exploits the inherent
parallelism of neural computation.

\begin{figure}[!t]
\begin{center}
\includegraphics[width=\columnwidth]{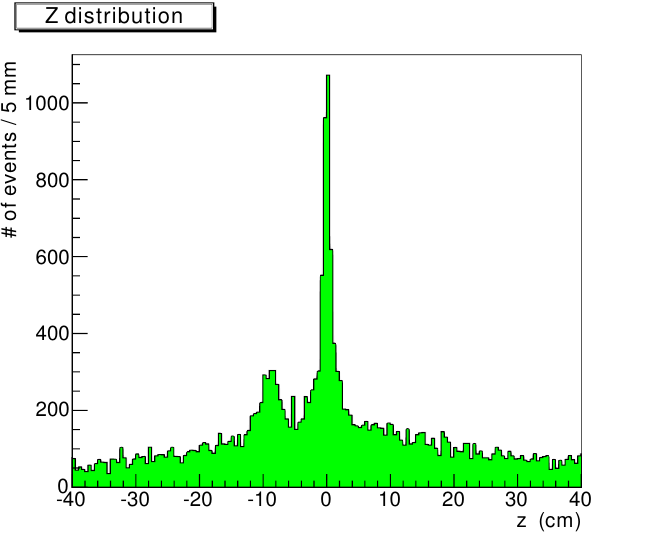}
\caption{Distribution of the $z$-position of reconstructed
  vertices in
  Belle~\cite{Belle2TDR}. The $z$-axis is parallel to the beam.
  The peak at $z=0\,\mathrm{cm}$
  corresponds to signal decays, the wide background is due to
  the Touschek effect and beam-gas interactions. The second peak
  at $z\approx-10\,\mathrm{cm}$ 
  is an artifact of the bunch structure of the beam.}
\label{zdist} 
\end{center}
\end{figure}

\IEEEpubidadjcol

Belle\,II~\cite{Belle2TDR} is an experiment at the asymmetric
electron-positron collider SuperKEKB~\cite{superkekb, superkekb2}, which is
currently under construction 
at the KEK laboratory in Tsukuba, Japan. Belle\,II is an upgrade of
the Belle experiment~\cite{BelleDetector}, which was instrumental in
exploring Charge Parity (CP) violation 
in the $B$ meson system. The B factories Belle and 
BaBar~\cite{BaBarDetector} jointly provided the experimental results 
confirming the Cabibbo Kobayashi Maskawa mechanism as the main source of CP 
violation in the standard model~\cite{Belle2TDR}. 
The success of this program has led to the rapid
approval of an upgrade for both the detector and the KEKB
collider. The new facility, Belle\,II at SuperKEKB, aims for an
amount of recorded physics events fifty times larger than in Belle,
{i.e.}, 50\,$\mathrm{ab}^{-1}$ at the energy of the $\Upsilon(4S)$ resonance.

The $\Upsilon(4S)$ decays with more than 95\,\% probability
into a pair of $B\bar{B}$ mesons. Therefore, the violation of CP
symmetry can be studied in this system with high precision.
In addition, the physics program of Belle~II includes heavy flavor
physics, $\tau$-lepton physics, $c$-quark physics, spectroscopy
and pure electroweak processes~\cite{superkekb, superkekb2}.
Belle~II will allow an indirect search for New Physics (NP) by precision 
measurements of flavour physics decay channels and thus complements the direct 
search for NP at high energy colliders~\cite{Belle2TDR}.

To reach the desired amount of physics events,
the instantaneous luminosity will exceed ${\cal L} = 8
\times 10^{35}\,\mathrm{cm}^{-2}\,\mathrm{s}^{-1}$, 40 times larger
than the world record achieved by KEKB. The pure physics rate at this luminosity is around $10\,\mathrm{kHz}$.
An unfortunate side effect of the high luminosity is a much higher
level of machine background, dominated by Touschek scattering~\cite{Touschek,Piwinski}.
This produces 
a high rate of undesirable background events with vertices outside of
the nominal interaction region, where the physically
interesting reactions from the $e^+e^-$ collisions are produced. As an
illustration, Fig.~\ref{zdist} shows the distribution of interaction vertices
in the beam direction $z$ as measured in Belle. The wide background
around the narrow peak at $z=0\,\mathrm{cm}$, marking the interaction point, is
clearly visible. Since Belle had no fast 
detection of the $z$-vertex at the L1 trigger, these events
could not be rejected there, leading to a signal to noise
ratio below $1:10$. 
Note that the
level of this background is expected to be much higher at SuperKEKB, 
due to the increased beam currents and the new beam optics (nano-beam 
option)~\cite{Belle2TDR}.

The showcase of the proposed project is the reduction of the 
Touschek scattering 
background by using track information from the CDC of Belle\,II. This requires a precise estimate (on the
order of 1 to 2\,cm) of the event $z$-vertex in real time,
sufficiently fast for the L1 trigger. 

\section{The Belle\,II Detector}
We give here a short description of the Belle\,II detector now under
construction, emphasizing the foreseen tracking and trigger
systems. More details can be found in~\cite{Belle2TDR}. 
\subsection{Overview}
As the range of center-of-mass energies of SuperKEKB is basically the
same as for KEKB, changing only mildly the energy asymmetry, the
physics requirements of Belle\,II are very similar to those of
Belle. This allows re-using the larger structures such as the
electromagnetic calorimeter, the solenoid, and the instrumented iron
return yoke. However, the much more demanding physics rate and
background requirements ask for entirely new tracking and particle
identification systems as well as for faster signal processing in the outer
detectors.  

The Belle\,II detector, matching these demands, consists of the following components.
The beam pipe is surrounded by a new vertex
detector (VXD).  The VXD consists of the pixel detector (PXD) with two layers
of pixel sensors in DEPFET technology, and the silicon vertex
detector (SVD) with four layers of double-sided strip sensors. 
The VXD provides a precise (offline) determination of the decay and
interaction vertices. 
The main tracking device is the large new CDC with axial and stereo wires.
Outside of the CDC, a new particle 
identification system (PID) is installed (TOP in the barrel, ARICH in the forward region).
The PID system is surrounded by the electromagnetic calorimeter (ECL),
followed by the Belle superconducting coil, producing a solenoidal
field of 1.5\,T. The instrumented
magnetic flux return yoke of Belle completes the outer dimensions of
the Belle\,II detector. 

\subsection{The Central Drift Chamber}
The new CDC contains about 50.000 sense and field wires, defining  
drift cells of size about 2\,cm in a cylindrical volume 
with inner radius of $r\approx16\,\mathrm{cm}$ 
and outer radius of $r\approx113\,\mathrm{cm}$.
The sense wires are arranged in layers, where 6 or 8 adjacent layers
are combined in a so-called SuperLayer (SL). 
The outer eight SL consist of 6 layers with 160 to 384 wires each. The  
innermost SL has 8 layers, each with 160 wires in smaller
(half-size) drift cells to cope with the 
increasing occupancy towards smaller radii.

The SL alternate between axial (A) orientation, aligned
with the solenoidal magnetic field ($z$-axis), and stereo (U,~V)
orientation. The stereo wires are skewed by an angle between 45.4 and
74\,$\mathrm{mrad}$ in positive and negative direction.   
The direction changes sign between U and V layers, with a
total SL configuration of AUAVAUAVA.

By combining the information of axial and stereo
wires (the space point resolution of the drift chamber is about
100\,$\mu\mathrm{m}$), it is possible to reconstruct a full 3D track. %In 

\subsection{The Trigger System}
Similar to Belle~\cite{BelleDetector}, the main first level triggers of Belle~II are based
on the CDC and the ECL. Triggers
will also be derived from the outer systems.
The Global Decision Logic, which combines the results of all
subtriggers and makes the final trigger decision,
is still under design, but a
trigger will generally require one or more charged tracks.

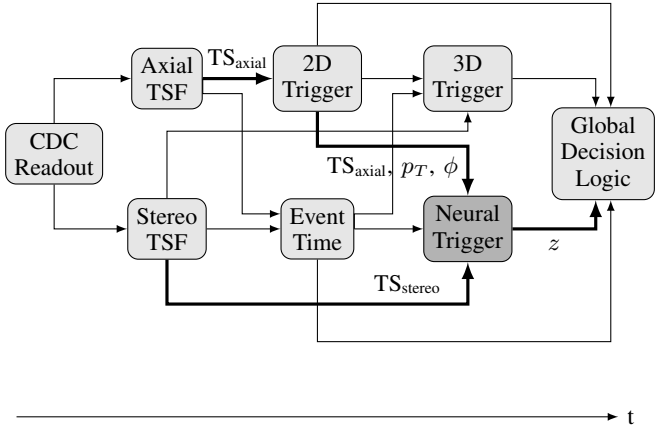
\begin{figure}[!t]
\centering
\begin{tikzpicture}
	\tikzstyle{every node}=[font=\small]
	\draw[-latex] (-2,-2.5) -- (6,-2.5) node[right] {t};
	%\node[draw,rounded corners,fill=black!10,align=center] (cdc) at (-5.5,1) {CDC};
	%\node[draw,rounded corners,fill=black!10,align=center] (digitizer) at (-3.5,1) {Digitizer};
	\node[draw,rounded corners,fill=black!10,align=center] (merger) at (-1.5,1) {CDC\\Readout};
	\node[draw,rounded corners,fill=black!10,align=center] (axial) at (0,2) {Axial\\TSF};
	\node[draw,rounded corners,fill=black!10,align=center] (stereo) at (0,0) {Stereo\\TSF};
	\node[draw,rounded corners,fill=black!10,align=center] (2D) at (2,2) {2D\\Trigger};
	\node[draw,rounded corners,fill=black!10,align=center] (3D) at (4,2) {3D\\Trigger};
	\node[draw,rounded corners,fill=black!30,align=center] (NN) at (4,0) {Neural\\Trigger};
	\node[draw,rounded corners,fill=black!10,align=center] (time) at (2,0) {Event\\Time};
	\node[draw,rounded corners,fill=black!10,align=center] (gdl) at (5.8,1) {Global\\Decision\\Logic};
	%\draw[-latex] (cdc) -- (digitizer);
	%\draw[-latex] (digitizer) -- (merger);
	\draw[-latex] (merger) |- (axial);
	\draw[-latex] (merger) |- (stereo);
	\draw[-latex,line width=1.2pt] (axial) -- node[sloped,above] {$\text{TS}_\text{axial}$} (2D);
	\coordinate (axialtime) at (1,0.2);
	\draw[-latex] (axial.east) |- (1,1.8) -- (axialtime) -- (axialtime -| time.west);
	\draw[-latex] (stereo) -- (time);
	\draw[-latex] (stereo) -- (0,1.3) -| (3D);
	\draw[-latex,line width=1.2pt] (stereo) -- (0,-1) -| node[sloped,above,pos=0.4] {$\text{TS}_\text{stereo}$} (NN);
	\draw[-latex] (2D) -- (3D);
	\draw[-latex,line width=1.2pt] (2D) -- (2,1.1) -| node[sloped,below,near start] {$\text{TS}_\text{axial}$, $p_T$, $\phi$} (NN);
	\coordinate (2Dgdl) at (5.9,3);
	\draw[-latex] (2D) |- (2Dgdl) -- (2Dgdl |- gdl.north);
	\coordinate (time3D) at (3,1.8);
	\draw[-latex] (time.east) |- (3,0.2) -- (time3D) -- (time3D -| 3D.west);
	\draw[-latex] (time) -- (NN);
	\coordinate (timegdl) at (5.9,-1.5);
	\draw[-latex] (time) |- (timegdl) -- (timegdl |- gdl.south);
	\coordinate (3Dgdl) at (5.7,2);
	\draw[-latex] (3D) -- (3Dgdl) -- (3Dgdl |- gdl.north);
	\coordinate (NNgdl) at (5.7,0);
	\draw[-latex,line width=1.2pt] (NN) -- node[sloped,below] {$z$} (NNgdl) -- (NNgdl |- gdl.south);
	%\draw (-2,0) node[draw,rounded corners,fill=black!10,align=center] (cdc) {CDC};
\end{tikzpicture}
\caption{Schematic of the CDC trigger. The
proposed neural $z$-vertex trigger takes input from the track
segment finders for the stereo superlayers and from the 2D
trigger. It will run in parallel to the standard 3D track trigger.}
\label{fig:signalflow}
\end{figure}

A schematic view of the track trigger is shown in Fig.~\ref{fig:signalflow}. 
The signals from the 9 SL with 14336 sense wires in total are
first read out in parallel 
in the front-end electronics boards, where they are digitized by
$10\,\mathrm{bit}$ $63.5\,\mathrm{MHz}$ ADCs, producing a data
rate coming
out of the CDC of $1784\,\mathrm{Gbps}$.   
These digital signals are then multiplexed by
merger boards. 

\begin{figure}[!t]
\centering
\includegraphics[width=\columnwidth]{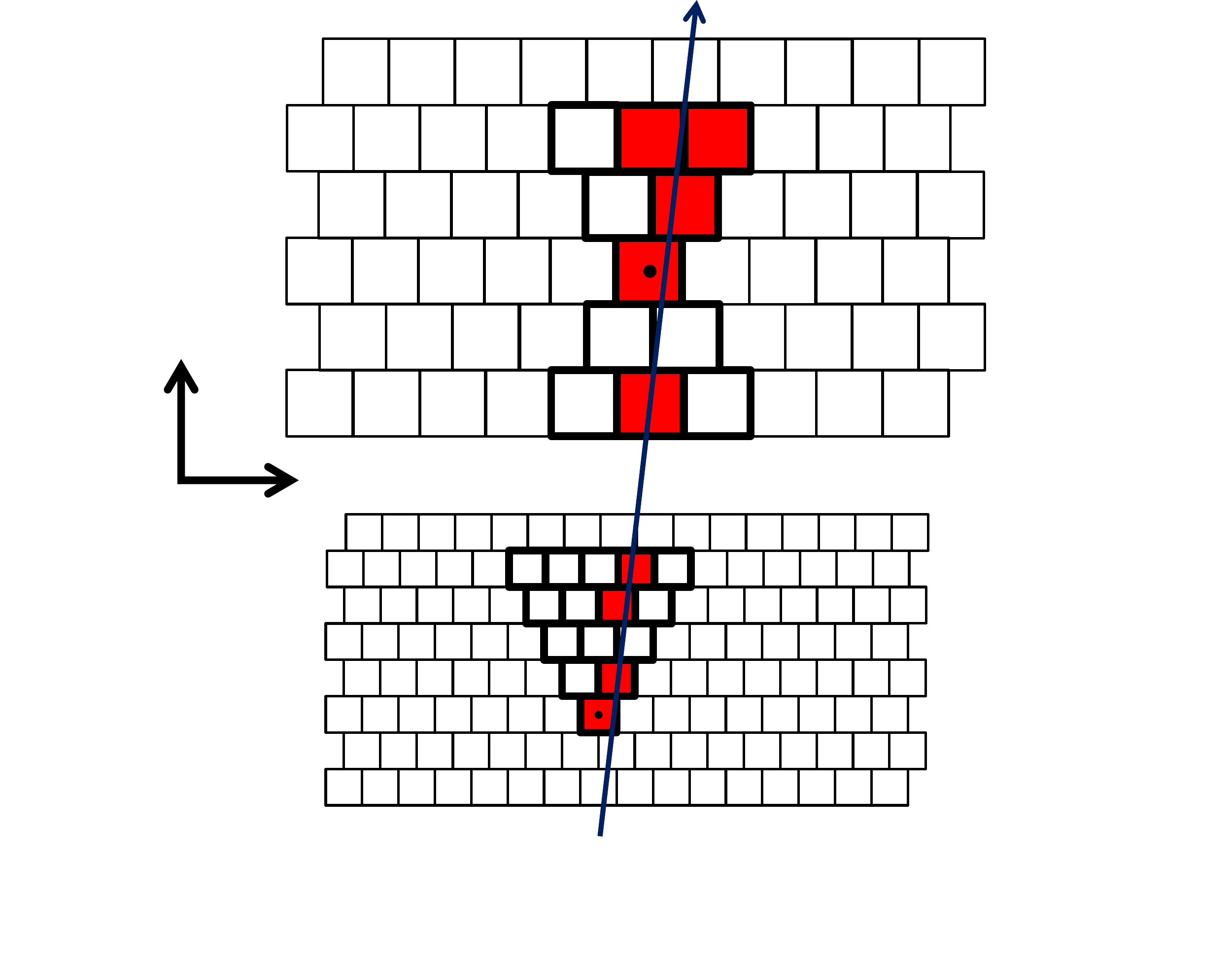}
\caption{A TS consisting of 5 layers in an SL. The
priority wire is indicated by a dot. Top: one of SL 2--9 with the
hourglass-shape, bottom: the innermost SL~1 with a different shape
and smaller drift cell size.}
\label{fig:hourglass}
\end{figure}

The strategy of the present CDC track trigger is based on so-called
Track Segments (TS), which are  produced 
for each of the 9 SL by the Track Segment Finder (TSF). A
TS is 
topologically defined by an ``hour glass'' shaped arrangement of 5
given layers within an SL
with a ``priority wire'' in the
central layer, and 2 wires in each of the 
adjacent layers, followed by 3 wires in each of the two layers
further out (see Fig.~\ref{fig:hourglass}). Within an SL many such TS can be
formed to cover the full azimuthal range. A total of
2336 TS are pre-defined in the entire CDC. The TSF in the CDC trigger logic combines the 
information in each of the hour glass regions and produces a TS if at least 4
wires in different layers have a hit. The TSF then
transmits the TS number (id) 
and the drift time of the priority wire for further processing (see
``2D Trigger'' and ``3D Trigger'' in Fig.~\ref{fig:signalflow}). 
With 8 bit precision the drift times from the TSF have a 2\,ns resolution and thus refer to a time interval of 512\,ns.

It is important to note that these drift times are not absolute
drift times with respect to the a priori unknown event time, but
only relative drift times contained in the current time window of
512\,ns, i.e. drift times from the TSF have a random offset. 
An additional event timing module, operating in parallel to the 2D
finder, will provide an estimate of the timing based on the fastest
TS hits out of the active TS in the event. This event time
estimate can be used to compensate the random offset for the
following trigger components. To make use of the event time in the
2D prediction, the second part of the 2D trigger (the 2D fitter)
will run on the boards of the 3D trigger and the neural trigger.

The 2D trigger follows the strategy of the previous Belle CDC
trigger, combining TS from the SL with axial wire
orientation to provide 2D tracks in 
$r-\phi$ space \cite{iwasaki}. At first a Hough finder separates
the tracks and provides rough estimates for $p_T$ and $\phi$ based
only on the TS ids of the axial SL, followed by a 2D fit where the
drift times of the axial layers are included in order to achieve a
higher precision. Following the idea of the BaBar trigger
upgrade~\cite{BaBarTRGUpgrade}, the 3D trigger is designed to provide 3D
tracks in order to determine the $z$-vertex of the
event and to reject events not coming from the primary
vertex. Combining hits in the stereo SL with the 2D tracks, the 
3D trigger obtains the $z$-coordinate of the individual hits and
performs a linear regression to find the polar angle and the $z$-vertex. Further details on the L1 
trigger system in Belle~II can be found in~\cite{iwasaki, Belle2TDR} 
and details on the adopted trigger scheme of Belle 
can be found in~\cite{BelleDetector}.

The neural trigger is operating in parallel to the 3D trigger 
and fulfills the same task, but with a new multivariate algorithm,
providing an independent estimate for the $z$-vertex. 
Since neural networks are general function approximators capable of 
learning nonlinear dependencies, they enable a stable
3D track reconstruction also in the presence of noise~\cite{FAMA}
and inhomogeneities 
in the electrical (drift) and magnetic (solenoid) fields. Compared to statistical optimal tracking 
methods used in the offline analysis (e.g. Kalman filter), which are too slow for 
the use in an online trigger, neural networks provide a good tradeoff between 
execution speed and prediction accuracy.
The outputs of the 2D/3D triggers 
and the neural network are finally fed into the Global
Decision Logic (see
Fig.~\ref{fig:signalflow}).

\subsection{Planned hardware solution}
\begin{figure}[!t]
\def\svgwidth{\columnwidth}
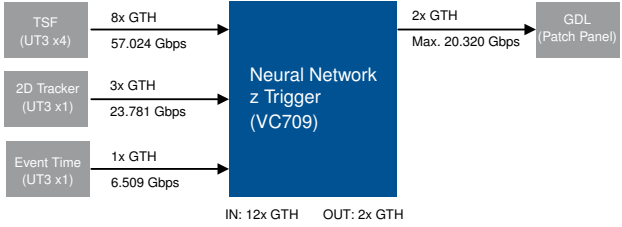
\caption{Interconnection of the neural board to the other
subsystems. One neural board is shown in the figure, together with
the bandwidths of data connections to other trigger components.
Four neural boards are used in total.}
\label{fig:nntt2subsys}
\end{figure}

The data transmission between the individual subsystems of the CDC
trigger is carried out via optical links with high speed serial
transceivers. Fig.~\ref{fig:nntt2subsys} illustrates how the
neural network trigger is connected to other trigger components.
The hardware system of the neural trigger consists of four FPGA
boards, each covering $180^\circ$ in the $r-\phi$ plane and having
$90^\circ$ overlap with its neighbors. Each neural board is
connected to four stereo TSF boards, one 2D trigger board and one
event timing board. Based on the system real-time requirements and
the resulting throughput calculation, the TSF boards deliver an
aggregate data bandwidth of $57.024\,\mathrm{Gbps}$, while the
data bandwidths coming from the 2D trigger board and the
event timing board add up to $23.781\,\mathrm{Gbps}$ and
$6.509\,\mathrm{Gbps}$, respectively. Taking into account that a
single GTH lane has an actual data rate of $10.160\,\mathrm{Gbps}$,
each neural board will require 12 GTH channels for data input (2 GTH per TSF board, 3 GTH for the 2D trigger board and 1 GTH for the event timing board).
Two additional GTH channels are reserved for data output to the GDL.            

The envisaged hardware solution for the CDC $z$-vertex trigger is
based on the Xilinx FPGA VC709 Connectivity Kit. 
It is equipped with a Xilinx Virtex-7 XC7VX690T FPGA with 3600
dedicated hardware multipliers, which is more than four times
the DSP resources available on Virtex-6, its previous generation.
This architecture is essential for our method, which requires
realtime critical, fully parallelized implementation of neural
networks or alternative multivariate methods. 
Furthermore, this platform features massive high-speed serial I/O
capability. There are 4 GTH lanes readily available on the main
board. By connecting the Xilinx \mbox{FM-S18} daughterboard to the FMC
connector of the main board, another 10 GTH lanes can be added to
the system. With 14 GTH lanes in total, it delivers sufficient
I/O bandwidth to communicate with the other trigger components in
Belle~II which are implemented on UT3 trigger
boards~\cite{Belle2TDR} developed at KEK.  
Additionally, it has $8\,\mathrm{GB}$ $1600\,\mathrm{MTs}$ DDR3
memory, which
is necessary for the storage and rapid retrieval of the neural network
parameters. Extrapolating from our preliminary study with narrow
sectors, about $10^9$ parameters will be needed ($10^6$ MLPs with
$10^3$ weights each).

This platform is currently the only solution on the market which
can fulfill all the requirements on the I/O throughput for the
proposed trigger. Having an off-the-shelf platform with rapid
prototyping capability enables us to verify and improve our
algorithms without excessive amount of investment usually required
by customized hardware designs.

\section{MLP Prediction}
We introduce here the operation principles of the MLP and its
application to the $z$-vertex prediction. Each charged track in
the detector crosses the magnetic field (1.5\,T) parallel to the
$z$-direction and can therefore be parametrized by a helix
$(p_T, \phi, d, \theta, z)$, where $p_T$ is the transverse
momentum, $\phi$ the azimuthal angle at the vertex position, $d$
is the distance of the track to the beam line in the
transverse plane ($r - \phi$), $\theta$ is the polar angle with respect to the $z$-axis
and $z$ is the position along the beam line. The vertex is
considered to be the position of closest approach of the helix to
the $z$-axis. The displacement $d$ of background tracks is
neglected in our studies, because the main contribution of the expected 
background tracks is beam induced~\cite{iwasaki, Belle2TDR} 
and thus has $d=0\,\mathrm{cm}$. In addition, background tracks with large 
displacements in $d$ can be identified and rejected solely by the 2D trigger.

We will show experimental results of MLPs specialized to sectors
in the phase space $(p_T, \phi, \theta, z)$ and trained with Monte
Carlo (MC) data, demonstrating the high resolution that can be
achieved on the $z$-vertex and confirming results from preliminary studies~\cite{SSDA}.

%\IEEEPARstart{M}{achine} Learning Introduction
\subsection{MLP structure}
\begin{figure}[!t]
\centering
\begin{tikzpicture}
  \node[ball color=blue!40,circle,minimum size=0.5cm] (out) at (2.8,0) {};
  \draw[-latex] (out) -- node[above] {$z$} (3.6,0);
  \node (morehidden) at (1.4,-0.6) {$\vdots$};
  \node (moreinput) at (0,0.2) {$\vdots$};
  \node[ball color=black!100,circle,minimum size=0.5cm] (biashidden) at (1.4,-2.4) {};
  \node[ball color=black!100,circle,minimum size=0.5cm] (biasinput) at (0,-1.6) {};
  \draw[-latex] (biashidden) -- (out);
  \node[ball color=blue!40,circle,minimum size=0.5cm] (hidden) at (1.4,1.6) {};
  \draw[-latex] (hidden) -- node[right,pos=0.3]{$w_{kj}$} (out);
  \draw[-latex] (biasinput) -- (hidden);
  \node[ball color=blue!40,circle,minimum size=0.5cm] (in) at (0,0.8) {};
  \draw[-latex] (in) -- node[above,pos=0.3]{$w_{ji}$} (hidden);
  \node[ball color=blue!40,circle,minimum size=0.5cm] (in) at (0,-0.6) {};
  \draw[-latex] (in) -- (hidden);
  \foreach \hidden in {0.8,0,-1.4}
  {
    \node[ball color=blue!40,circle,minimum size=0.5cm] (hidden) at (1.4,\hidden) {};
    \draw[-latex] (hidden) -- (out);
    \draw[-latex] (biasinput) -- (hidden);
    \foreach \inputnode in {0.8,-0.6}
    {
      \node[ball color=blue!40,circle,minimum size=0.5cm] (in) at (0,\inputnode) {};
      \draw[-latex] (in) -- (hidden);
      \draw[-latex] (-1,\inputnode) -- node[above] {$x_i$} (in);
    }
  }
  \draw (0,2.5) node[align=center] {input\\layer};
  \draw (1.4,2.5) node[align=center] {hidden\\layer};
  \draw (2.8,2.5) node[align=center] {output\\layer};
\end{tikzpicture}

\caption{Structure of an MLP with one hidden layer. In the nodes
the weighted sum of the input values $x_i$ is evaluated by the
activation function to produce one output value. The weight
matrices for the connections are $w_{ji}$ for the input $x_i$ to
the hidden layer and $w_{kj}$ for the hidden to the output layer.
The black nodes denote the constant bias nodes represented by
$w_{j0}$ and $w_{k0}$. We scale all input values to the interval
$x \in [-1, 1]$.}
\label{mlpgraph}
\end{figure}
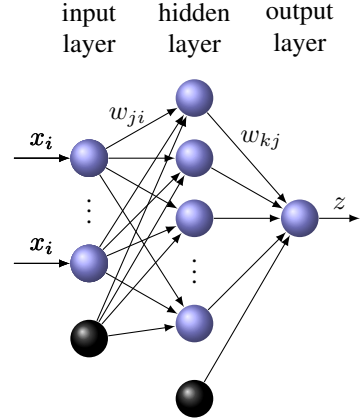

Neural networks are a biologically
inspired machine learning model. The MLP is a universal function
approximator based on the structure of a directed acyclic graph
(see Fig.~\ref{mlpgraph}).
The intrinsically parallel nature of MLPs is
perfectly suited for an implementation on parallel hardware
like FPGAs or GPUs. For the 3-layer MLP with one hidden layer, it
has been proven that it can approximate 
any continuous function to any required precision, given a hidden
layer of sufficient size and a non-constant, non-linear and
bounded activation function~\cite{perceptron-hypothesis}. Each
neuron computes the weighted sum over its input values and
evaluates it with an activation function

\begin{equation}
y_j = f \left( \displaystyle \sum_{i=0}^N w_{ji} \cdot x_i \right) 
= f ( w_{ji} \cdot x_i )
\end{equation}
where the sum in the last term is implicit for double indices.
Up to the activation function the calculation for one layer
corresponds to a matrix multiplication and can be similarly
parallelized, i.e. all neurons are calculated in parallel.
The weights $w_{ji}$ contain the information of the calculated
function. $y_j$ is the output value of neuron $j$, the input
vector $x_i$ has a constant $x_0 = 1$ to represent a constant bias
and the $x_i, i \in [1 .. N] $ are the input values.  The
activation function $f$ is in our case the hyperbolic tangent. The
complete function computed by an MLP with hidden nodes
$j$ at the output node $k$ is

\begin{equation}
z_k = f\left( w_{kj} \cdot f ( w_{ji} \cdot x_i ) \right)
\end{equation}
where $z_k$ are the $k$ outputs of the network, $w_{ji}$ and
$w_{kj}$ are the weight matrices connecting the input with the
hidden layer ($w_{ji}$) and connecting the hidden with the output
layer ($w_{kj}$). The summing over double indices is again
implicit. In our setup all input and output values of the MLP are
scaled within $[-1,1]$.

\subsection{MLP training}
The training of the MLP is based on a cost function $E$, the mean
squared error, applied to the MLP output

\begin{equation}
E = \displaystyle \sum^{N_\text{train}}_{i=1} (z_i - t_i)^2
\label{MSE}
\end{equation}
where $z_i - t_i$ is the deviation of the network output $z_i$
from the true value $t_i$ for the training event $i$ and
$N_\text{train}$ is the number of training events. In the
supervised learning scheme this cost function is minimized by
iteratively adjusting the weights in the network proportional to
the derivatives

\begin{equation}
\Delta w_{mn} \propto \frac{\partial E}{\partial w_{mn}}
\label{MSEderiv}
\end{equation}
Our network training is performed using the iRPROP$^-$
algorithm~\cite{igel-propose}, an improved variant of the
RPROP~\cite{rprop-original} backpropagation algorithm. This
algorithm has been demonstrated to be more effective than the
classical backpropagation~\cite{igel-evaluate}, because the
magnitude of the weight update is independent of the magnitude of
the derivative of the cost functions and only dependent on the
dynamics of the past weight updates. The effect is a faster
convergence to a minimum of the cost function with the same minima
found as with classical backpropagation~\cite{igel-evaluate}.

The training procedure can be parallelized by so called ``pattern
parallel training'', where parallelization
is achieved by splitting the set of training
patterns over several threads~\cite{pattern-parallel}. This is possible
because the cost function in \eqref{MSE} is a simple sum over the
deviations from the target value for each training pattern. Since
all summands are independent from each other, they can be
calculated in parallel and then summed up. The same is true for
the gradient of the cost function and therefore for the weight
updates, which can again be decomposed into a sum of independent
terms, each depending on one training pattern.

\subsection{Sectorization of Input Data}
\label{sec:sectorization}
The prediction of the $z$-vertex value with an MLP is possible if
the remaining track parameters are known with a sufficient
accuracy and thus form a sector in phase space, defined by
intervals of the helix track parameters. An ``expert'' MLP is
specialized to a sector by selecting the ``relevant'' TS that are
needed to describe the sector accurately~\cite{SSDA}. Only hits from
these ``relevant'' TS are used as input for the MLP. This approach is
motivated by~\cite{adamix}, where an ensemble of local ``experts''
is combined with a selection procedure for the optimal local
``expert''.

This selection is a necessary preprocessing step to provide
a-priori knowledge to the network and to reduce the number of
inputs of the MLP to a manageable level. The number of selected
``relevant'' TS is much smaller than the total number of TS
(2336), typically about 1\% of the total. Furthermore, it is
mainly dominated by the sector sizes in $p_T$ and $\phi$, whereas
the sector sizes in $\theta$ and $z$ only weakly influence the
stereo layers. 

\begin{figure}[!t]
\centering
\def\svgwidth{0.9\columnwidth}
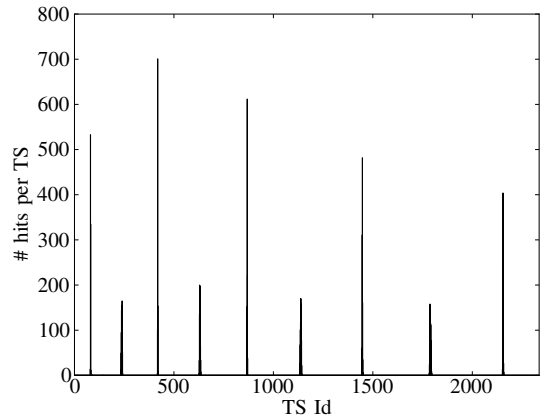
\caption{Selection of relevant TS in a sector of
$\phi \in [180^\circ,181^\circ]$, $\theta \in [35^\circ,123^\circ]$, $p_T \in [1.43,1.67]\,\mathrm{GeV}$,
$z \in [-50,50]\,\mathrm{cm}$. Each peak corresponds to an SL,
alternating between axial and stereo SL. The skewing angle of the
stereo SL leads to a broadening of the stereo peaks. The decrease
of the peaks towards the outer SL is due to the widening of the
sector towards larger radii.}
\label{tssector}
\end{figure}

For a defined sector in phase space, ``relevant'' TS can be
selected using MC datasets for the Belle~II detector. From these
datasets a histogram of the TS activity of events within a given
sector is generated (see Fig.~\ref{tssector}). Small TS ids
correspond to the inner layers and large TS ids to the outer SL
(e.g. TS id $\in [1, 160]$ is SL 1). The 9 peaks correspond to the
9 SL in the CDC and the selection of a small subset of
``relevant'' TS is obviously justified. On a per SL basis, it is
required that a minimum percentage of the hits in all events are
found in TS selected as ``relevant'' within the sector.

\subsection{MLP I/O}
Once the ``relevant'' TS for a sector are found, the hits in each
event need to be formatted as input for the MLP. While the number
of hits in different events can vary, the number of inputs for the
MLP needs to be fixed. Two different approaches are used in our
studies. The first is a topological input distribution, where each
input node corresponds to one relevant TS and the input values are
the drift times scaled to the interval $[-1, 1]$. Relevant TS that
have no hit in a given event are set to a default value
corresponding to the maximal drift time, i.e. the track is treated
as if it were far away from the TS.

The second approach uses a fixed number of two input nodes per SL,
where the first input corresponds to a drift time and the second
to a TS id, both scaled to the interval $[-1, 1]$. The numbering
of the TS within an SL is continuous and can therefore be
interpreted as a scaled azimuthal angle. In the rare case that an
event has 2 hits in the same SL, the fastest hit is used. In the
case that an SL has no hit due to limited efficiency, a default
value is used again.

The output of the MLP is also within $[-1,1]$ due to the
activation function applied in the output neuron. In order to get
a floating point prediction for a helix track parameter, the
output is scaled to the sector interval of that variable. While an
MLP can generally be trained to predict all track parameters, we
use it only to predict the $z$-vertex and the polar angle $\theta$
in our experiments.

\subsection{``Expert'' MLP results for small sectors}

\begin{figure}[!t]
%\vspace*{-\bigskipamount}
\begin{center}
\def\svgwidth{0.98\columnwidth}
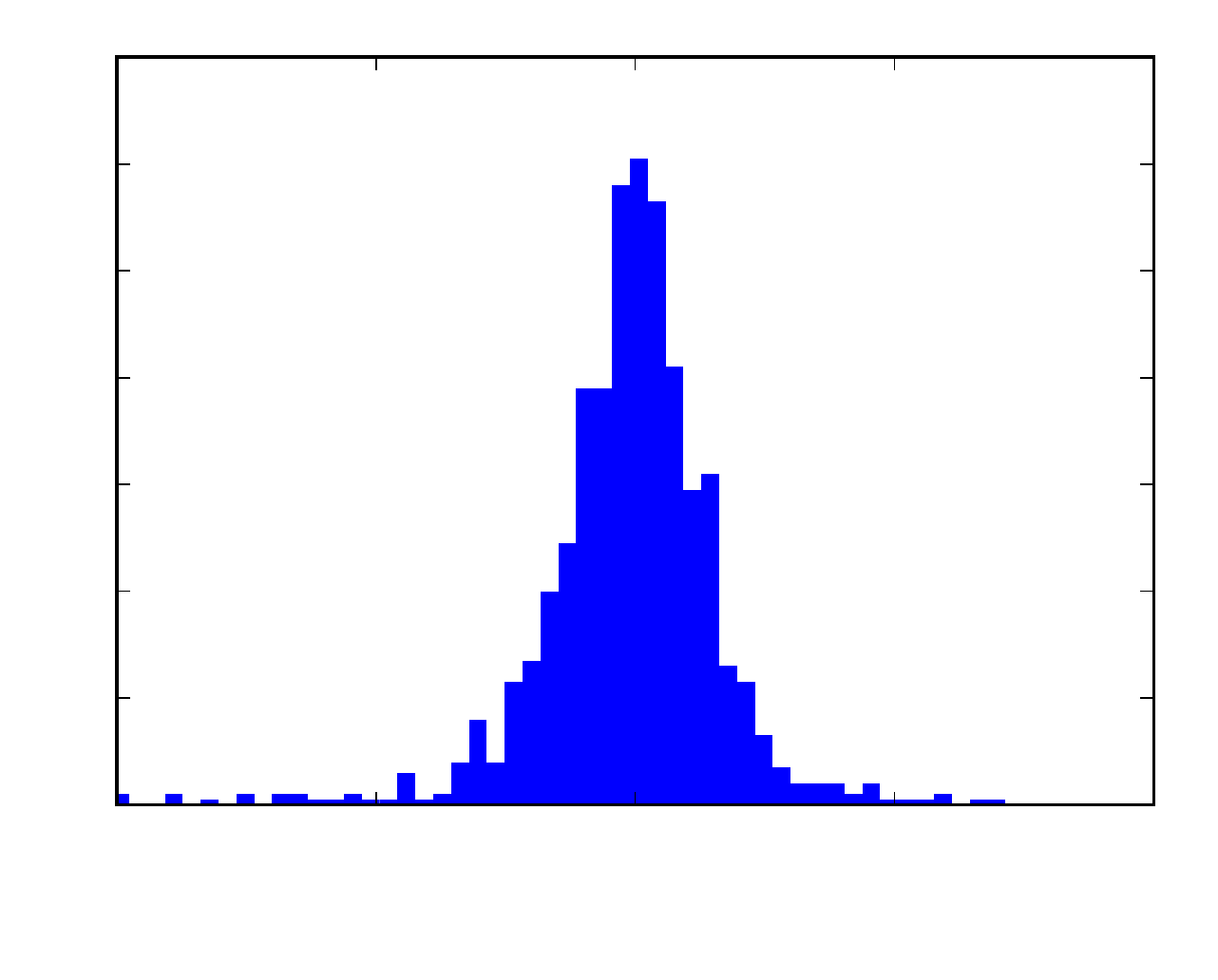
\def\svgwidth{0.98\columnwidth}
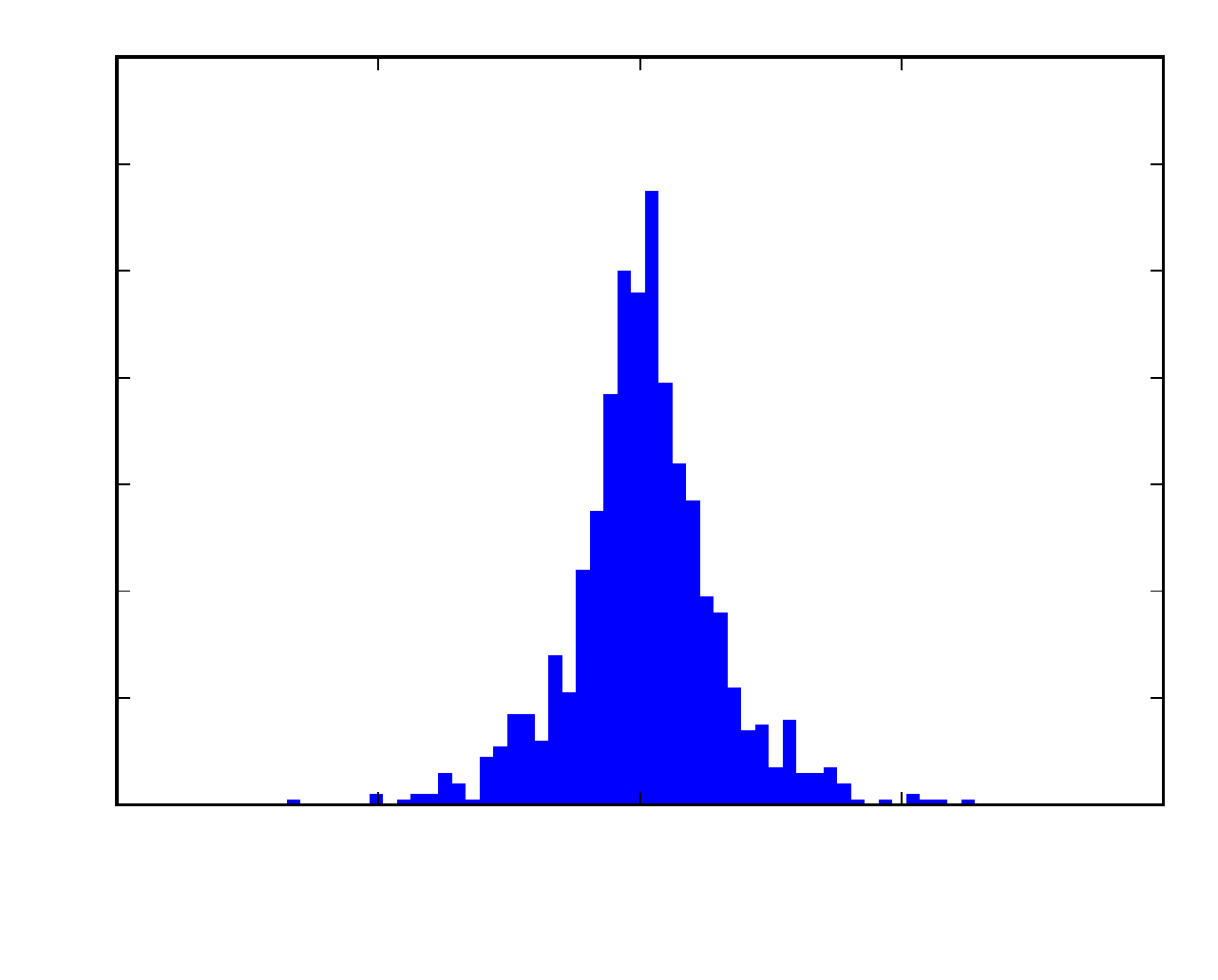

\vspace*{-\bigskipamount}
\caption{Achieved $z$-vertex resolution with MLPs 
  for charged single track events in a small sector constrained by the
  track parameters $\phi \in [180^\circ, 181^\circ]$,
  $\theta \in [56^\circ, 62^\circ]$, $z \in [-10,10]\,\mathrm{cm}$.
  Top plot: low momentum tracks with transverse
  momentum $p_T \in [0.500,0.513]\,\mathrm{GeV}$, bottom plot:
  high momentum tracks with $p_T \in [4.0,5.0]\,\mathrm{GeV}$.} 
\label{mlp-evidence}
\end{center}
\end{figure}

In order to show that a sufficiently precise $z$-vertex
reconstruction is possible at L1, we have tested the MLP approach
on single simulated muon tracks from narrow regions in polar and
azimuthal angle ($\theta$ and $\phi$), as well as in a limited
region of transverse momentum ($p_T$).
This setup allows to test single sectors without first training all $10^6$ MLPs.
In the actual trigger the charged tracks in an event will be
separated by the 2D finder, so the neural trigger will handle tracks one by one.
Training runs in various sectors showed that the required $z$-vertex
resolution can indeed be achieved by this method~\cite{SSDA}. The
results for two different $p_T$ regions for a single ``expert'' MLP (see
Fig.~\ref{mlp-evidence}) show that the $z$-resolutions with
$\mathrm{RMS} = 1.72\,\mathrm{cm}$ for the low $p_T$ case and
$\mathrm{RMS} = 1.44\,\mathrm{cm}$ for the high $p_T$ case are
well below our anticipated $2\,\mathrm{cm}$ resolution. A
fixed-point implementation suitable for FPGAs has also been
carried out and compared to the floating point reference design.
There is no evidence of deteriorated $z$-resolutions with
$\mathrm{RMS} = 1.73\,\mathrm{cm}$ for the low $p_T$ case and
$\mathrm{RMS} = 1.44\,\mathrm{cm}$ for the high $p_T$ case. 

A preliminary study with post synthesis simulation shows that the execution 
time of an MLP with 1260 nodes (20 input neurons, 60 hidden neurons, 1 output 
neuron) is $136\,\mathrm{ns}$. Based on the data throughput of the external memory 
on our hardware platform, the transfer time for network parameters is estimated to 
be $223\,\mathrm{ns}$. For the aforementioned network size and topology, the total 
latency of each processing cycle will be under $400\,\mathrm{ns}$. The latency is 
linearly proportional to the number of nodes in the MLP.   

Note that the sector size in $p_T$ needs to be much smaller for
the low $p_T$ case in order to achieve a comparable resolution for
both cases. This is because the geometrically relevant property of
the sector, namely the curvature of the track, is proportional to
$p_T^{-1}$. The two sectors were therefore chosen to have the same
size in $\Delta p_T^{-1}$ rather than in $\Delta p_T$.
In the experiment a Geant4 based detector simulation of the
Belle~II CDC is used. It includes the simulation of physics
effects due to material interactions with the inner detector
components, the non-linear $x-t$ relation~\cite{Belle2TDR} of
drift lengths to drift times due to inhomogeneities of the
electric field and the wire-sag effect caused by gravitation. With
the more realistic physics simulation, the $\mathrm{RMS}$
shown in Fig.~\ref{mlp-evidence} is about one third
higher than the results achieved in the preliminary studies \cite{SSDA, SSMA, FAMA},
where an idealized perfect detector was assumed using an older
version of the simulation software of the Belle~II detector.
The non-linear $x-t$ relation has the strongest effect on
the $\mathrm{RMS}$.

The generalization of this proof of concept to the full acceptance region
of the detector requires a pre-processing that can select the correct
sector for each track in the event. The correct selection of the
sector is discussed in the next section.

\section{Preprocessing}

We plan to run the full $z$-vertex trigger system for the Belle~II
detector as a two step prediction chain: At first the
preprocessing step provides the correct sector for an ``expert'' MLP.
Secondly, the ``expert'' MLP can provide the $z$-vertex as
demonstrated in the last section.
The experiments detailed in the following section show that the
MLP is also a promising candidate for the preprocessing when
combined with information from the 2D trigger. 

\subsection{Finding Sectors}
The finding of a sufficiently small sector is crucial for the
proper capability of an ``expert'' MLP to provide high accuracy
$z$-vertex predictions. The binning size in the phase space
variables is limited by the resolution of the sector finders. For
wrongly identified sectors the ``expert'' networks will fail
because the true result is out of range of their specification,
which makes the misclassification rate an interesting observable. 

As input to the neural network preprocessing, the 2D trigger
provides information on the number of tracks in the event and for
each track a prediction of $\phi$ and $p_T$, based on the active
TS and drift times in the axial layers. 
The appropriate sectors in the $p_T - \phi$ space will be provided 
by the 2D trigger. However, it is not yet clear whether the 2D trigger 
can yield the required resolution.

As stated before, the number of relevant TS depends mainly on
$\phi$ and $p_T$ (see section \ref{sec:sectorization}), so with
the information of the 2D trigger alone the number of relevant TS
can be reduced to a level suitable for the application of an MLP.
This first MLP is trained to predict $\theta$ and $z$. It does not
reach the final $z$-resolution of $2\,\mathrm{cm}$ as the
``expert'' MLP, but it is suitable as a preprocessing step. The
prediction of the ``preprocessing'' MLP for $z$ and $\theta$,
together with the estimate of $p_T$ and $\phi$ provided by the 2D
trigger, is then used to select an ``expert'' MLP which has limited
ranges in all track parameters.

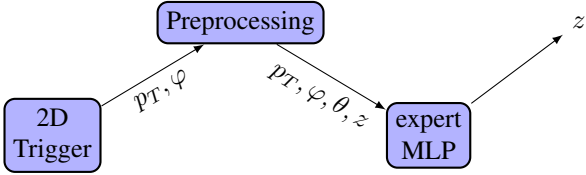
\begin{figure}
\centering
\begin{tikzpicture}
  \node[align = center, draw,rounded corners,fill=blue!30,line width=0.8pt] (2d) at (0,0){2D\\Trigger};   
  \draw (2.5,1.5) node[draw,rounded corners,fill=blue!30,line width=0.8pt] (run1) {Preprocessing};
  \draw (5,0) node[align=center, draw,rounded corners,fill=blue!30,line width=0.8pt] (run2) {expert \\ MLP};
  \draw (7,1.5) node (z) {$z$};
  \draw[-latex] (2d) -- node[sloped,below] {$p_T, \varphi$} (run1);
  \draw[-latex] (run1) -- node[sloped,below] {$p_T, \varphi, \theta, z$} (run2);
  \draw[-latex] (run2) -- (z);
\end{tikzpicture}
\caption{Data flow in the prediction chain: the $p_T$ and $\phi$
information obtained from the 2D Trigger is enriched with $\theta$
and $z$ information by the preprocessing. This allows selection
of an adequate ``expert'' MLP for the precise $z$-vertex prediction.
The MLP approach is also a promising candidate for the Preprocessing.}
\label{chainflow}

\end{figure}

The workflow of this two step prediction concept is illustrated in
Fig.~\ref{chainflow}. After one preprocessing step ($1^\text{st}$
round) the correct ``expert'' network ($2^\text{nd}$ round) is
chosen. Our concept can be extended to several steps if necessary.
In case the prediction of the ``preprocessing'' MLP is not precise
enough to select a small ``expert'' MLP, the preprocessing step
can again be divided into several steps with successively
decreasing sector sizes. In the following we describe an
experiment with 3 consecutive steps, where the final step achieved
the required $z$-resolution. Whether the required resolution can
also be achieved within only two steps will be determined by
future research.

\subsection{Experimental setup}
To demonstrate the capabilities of the prediction chain we use
again single simulated muon tracks restricted to a sector in phase
space. We started from a sector limited in 2D to
$\phi \in [180^\circ,181^\circ]$ and
$p_T \in [1.43,1.67]\,\mathrm{GeV}$. For the polar angle $\theta$
we chose the starting ranges $[35^\circ,123^\circ]$, which
corresponds to the region for which straight tracks coming from
the interaction point pass all 9 SL of the CDC. For $z$ the
starting range was chosen as $[-50,50]\,\mathrm{cm}$. Within these
ranges single tracks were simulated and used to train two MLPs,
one with topological input distribution, the other with TS ids as
additional input nodes. Both MLPs are trained to predict $z$ and
$\theta$. For the selection of the next sector the predictions of
both MLPs are averaged, since the combined prediction reaches a
better resolution than each MLP alone. The use of a mixture of
different ``experts'' specialized to the same phase space sector
is inspired by~\cite{gasen}, where a more elaborate combination
procedure is proposed (GASEN algorithm).

After testing the resolution of the first step, we define a set of
sectors for the second steps based on the measured resolution. For
the $z$-vertex the new ranges are determined such that 99\,\% of
the events from $z \in [-1,1]\,\mathrm{cm}$, i.e. possibly
interesting physics events, are predicted within the narrowed
ranges. Any events predicted outside of the new sector can be
safely rejected already in the first step. The $\theta$-resolution
is then tested with events from the narrowed $z$-ranges and the
$\mathrm{RMS}$ of the difference between true and predicted
$\theta$ is calculated. A set of sectors is defined with sector
sizes of ${6\cdot\mathrm{RMS}(\theta)}$, which corresponds to a
$3\sigma$ interval in both directions. In order to avoid binning
effects the sectorization is done with an overlap. With two
overlapping binnings, displaced by a half binsize relative to each
other, predictions close to the sector border find another sector
where they are close to the center.

\begin{figure}[!t]
\centering
\def\svgwidth{0.98\columnwidth}
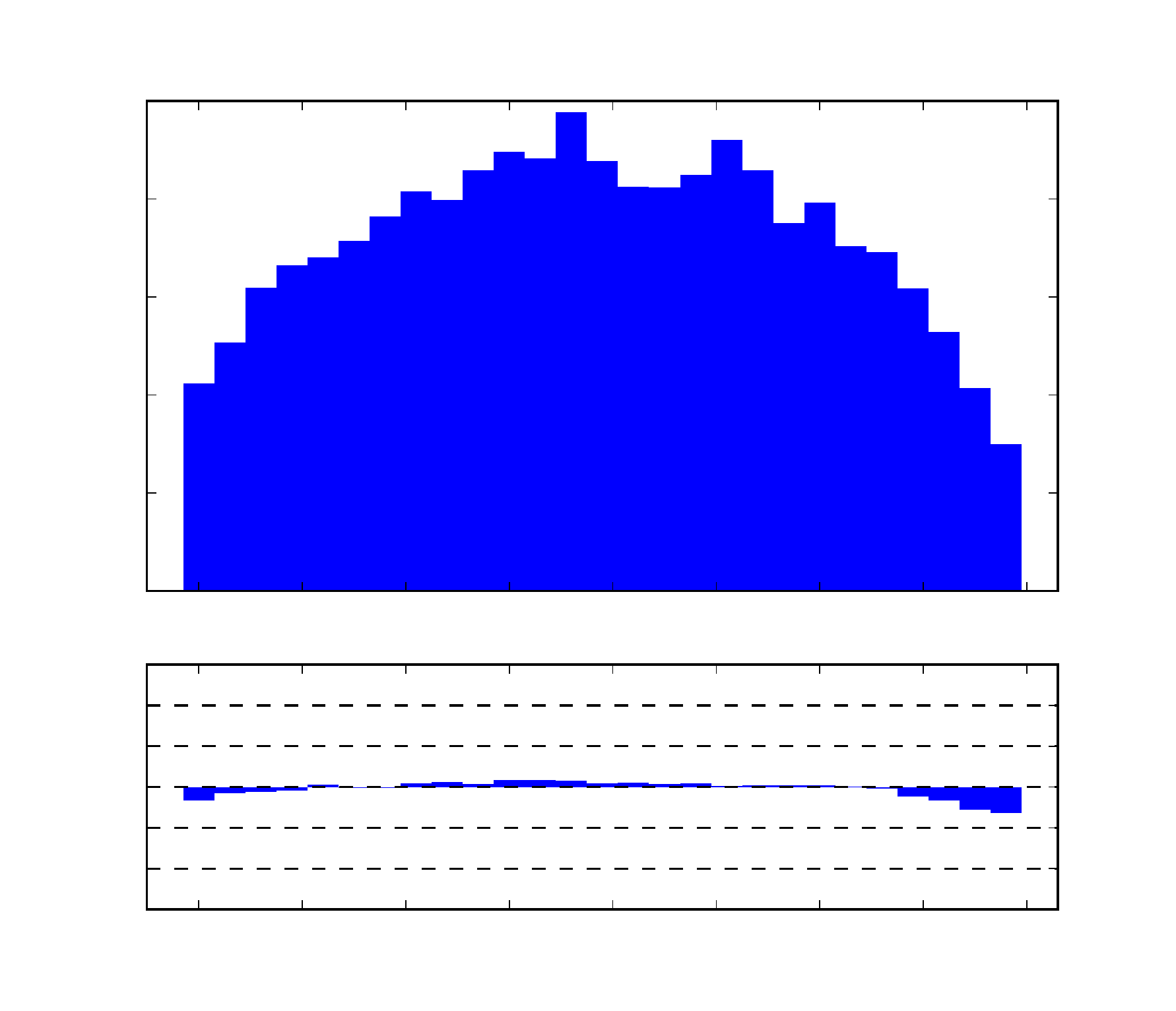
\def\svgwidth{0.98\columnwidth}
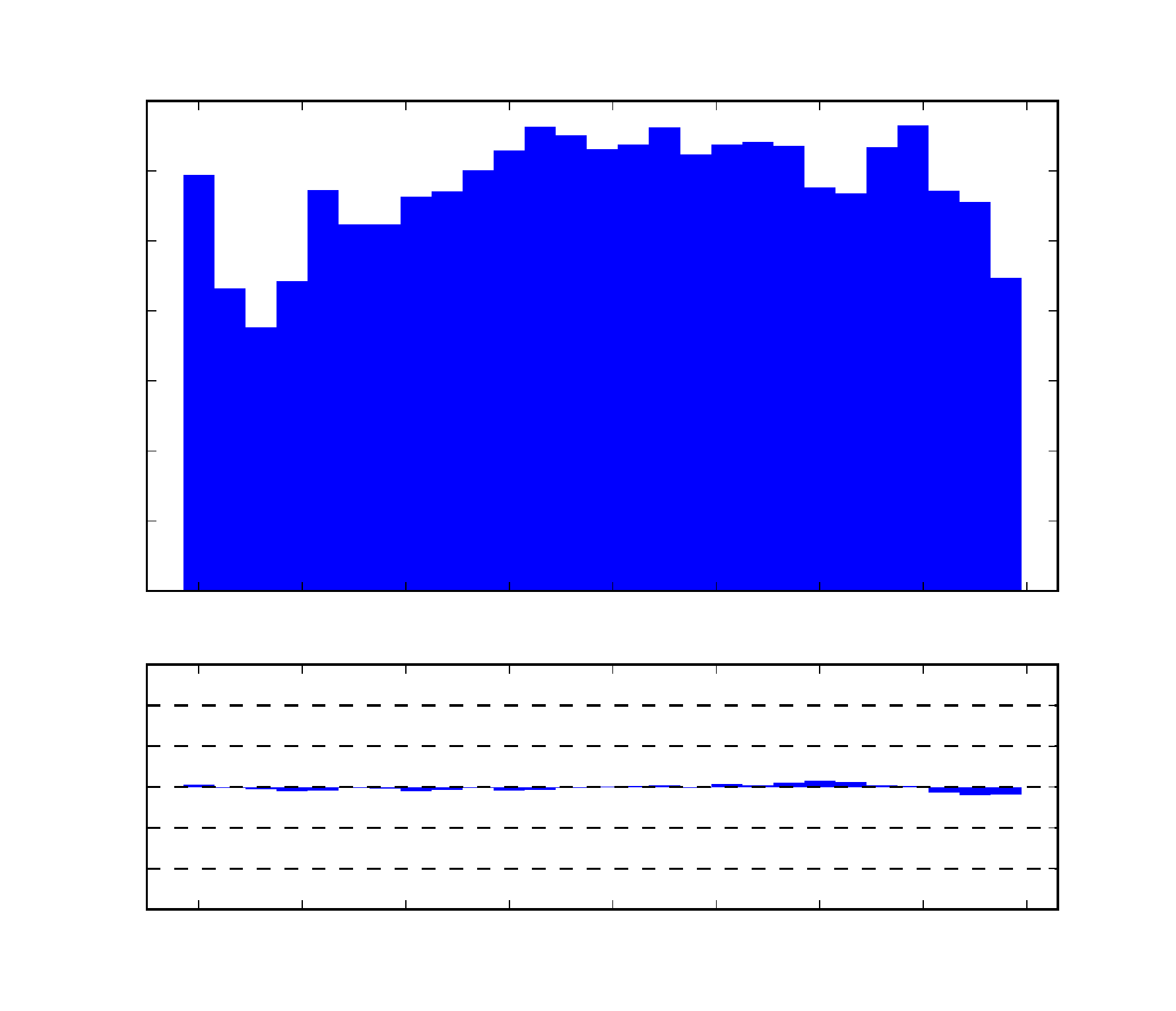
\caption{Resolution of $\theta$ depending on $\theta$ for an MLP
with topological input distribution (top) and an MLP with TS ids as
additional input nodes (bottom).}
\label{theta}
\end{figure}

The same procedure is repeated to define the sectors for the third
step. For each sector we train again one MLP with topological
input distribution and one with TS ids as additional input nodes
to predict $z$ and $\theta$. For the test of the second step the
MLPs of the first step are used to select the sector in $\theta$,
i.e. the measured resolution already includes events that were
predicted in the wrong sector. Finally the resolution of the third
step is measured, again using the first two steps to find the
correct sector.

\subsection{Prediction of the polar angle $\theta$}
The $\theta$-resolution of the preprocessing is of special
interest because the performance of the ``expert'' MLP depends
strongly on the sector size in $\theta$. The results for both MLPs
of the first step are shown in Fig.~\ref{theta}. The MLP with
topological input reaches a different resolution for different
$\theta$-regions, while the MLP with TS ids as additional inputs
does not depend strongly on $\theta$. On average, the topological
MLP reaches a resolution of $\mathrm{RMS}(\theta) = 5.8^\circ$,
the MLP with TS ids reaches a resolution of
$\mathrm{RMS}(\theta) = 5.2^\circ$ and the combined prediction
reaches a resolution of $\mathrm{RMS}(\theta) = 5.1^\circ$. For
the second step the sector size in $\theta$ is therefore set to
$6 \cdot \mathrm{RMS}(\theta) = 30.5^\circ$.

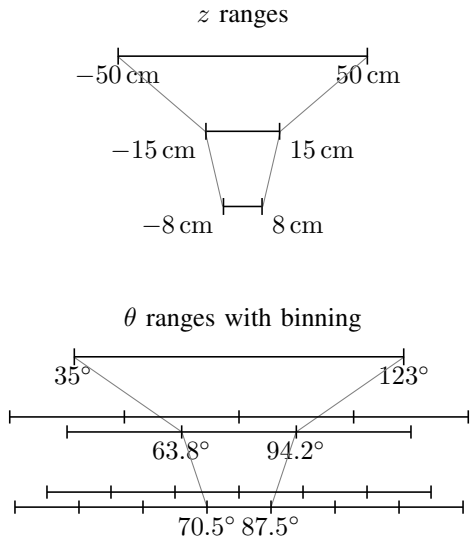
\begin{figure}
\centering
\begin{tikzpicture}
	\node (z) at (0,2.5) {$z$ ranges};
  \coordinate[label=below:$-50\,\mathrm{cm}$] (z0minus) at (-1.667,2);
  \coordinate[label=below:$50\,\mathrm{cm}$] (z0plus) at (1.667,2);
  \coordinate[label=below left:$-15\,\mathrm{cm}$] (z1minus) at (-0.5,1);
  \coordinate[label=below right:$15\,\mathrm{cm}$] (z1plus) at (0.5,1);
  \coordinate[label=below left:$-8\,\mathrm{cm}$] (z2minus) at (-0.267,0);
  \coordinate[label=below right:$8\,\mathrm{cm}$] (z2plus) at (0.267,0);
  \draw[color=black!50] (z0minus) -- (z1minus) -- (z2minus);
  \draw[color=black!50] (z0plus) -- (z1plus) -- (z2plus);
  \draw[|-|,semithick] (z0minus) -- (z0plus);
  \draw[|-|,semithick] (z1minus) -- (z1plus);
  \draw[|-|,semithick] (z2minus) -- (z2plus);
  
  \node (theta) at (0,-1.5) {$\theta$ ranges with binning};
  \coordinate[label=below:$35^\circ$] (theta0minus) at (-2.25,-2);
  \coordinate[label=below:$123^\circ$] (theta0plus) at (2.15,-2);
  \coordinate[label=below:$63.8^\circ$] (theta1minus) at (-0.81,-3);
  \coordinate[label=below:$94.2^\circ$] (theta1plus) at (0.71,-3);
  \coordinate[label=below:$70.5^\circ$] (theta2minus) at (-0.475,-4);
  \coordinate[label=below:$87.5^\circ$] (theta2plus) at (0.375,-4);
  \draw[color=black!50] (theta0minus) -- (theta1minus) -- (theta2minus);
  \draw[color=black!50] (theta0plus) -- (theta1plus) -- (theta2plus);

  \draw[|-|,semithick] ($- 0.05*(44,0) + 0.05*(79,0) + (-4,-2)$) -- ($0.05*(44,0) + 0.05*(79,0) + (-4,-2)$);
  \draw[decorate,decoration={ticks,segment length=1.524cm},semithick] ($0.05*30.48*(-1,0) - 0.05*(15.24,0) + 0.05*(79,0) + (-4,-3)$) -- ($0.05*30.48*(1,0) + 0.05*(15.24,0) + 0.05*(79,0) + (-4,-3)$);
  \draw[semithick] ($0.05*30.48*(-1,0) - 0.05*(15.24,0) + 0.05*(79,0) + (-4,-3)$) -- ($0.05*30.48*(1,0) + 0.05*(15.24,0) + 0.05*(79,0) + (-4,-3)$);
  \draw[decorate,decoration={ticks,segment length=1.524cm},semithick] ($0.05*30.48*(-1.5,0) - 0.05*(15.24,0) + 0.05*(79,0) + (-4,-2.8)$) -- ($0.05*30.48*(1.5,0) + 0.05*(15.24,0) + 0.05*(79,0) + (-4,-2.8)$);
  \draw[semithick] ($0.05*30.48*(-1.5,0) - 0.05*(15.24,0) + 0.05*(79,0) + (-4,-2.8)$) -- ($0.05*30.48*(1.5,0) + 0.05*(15.24,0) + 0.05*(79,0) + (-4,-2.8)$);
  \draw[decorate,decoration={ticks,segment length=0.851cm},semithick] ($0.05*17.02*(-3,0) - 0.05*(8.51,0) + 0.05*(79,0) + (-4,-4)$) -- ($0.05*17.02*(3,0) + 0.05*(8.51,0) + 0.05*(79,0) + (-4,-4)$);
  \draw[semithick] ($0.05*17.02*(-3,0) - 0.05*(8.51,0) + 0.05*(79,0) + (-4,-4)$) -- ($0.05*17.02*(3,0) + 0.05*(8.51,0) + 0.05*(79,0) + (-4,-4)$);
  \draw[decorate,decoration={ticks,segment length=0.851cm},semithick] ($0.05*17.02*(-2.5,0) - 0.05*(8.51,0) + 0.05*(79,0) + (-4,-3.8)$) -- ($0.05*17.02*(2.5,0) + 0.05*(8.51,0) + 0.05*(79,0) + (-4,-3.8)$);
  \draw[semithick] ($0.05*17.02*(-2.5,0) - 0.05*(8.51,0) + 0.05*(79,0) + (-4,-3.8)$) -- ($0.05*17.02*(2.5,0) + 0.05*(8.51,0) + 0.05*(79,0) + (-4,-3.8)$);
\end{tikzpicture}
\caption{Sector sizes in $z$ (top) and $\theta$ (bottom) for a prediction chain with 3 steps. The $z$-ranges are iteratively decreased. The $\theta$-ranges in step 2 and 3 are covered with overlapping sectors of decreasing width. The selected sector shown in the figure is an example for an event from $\theta \approx 80^\circ$.}
\label{fig:predictionchain}
\end{figure}

\subsection{Three step prediction chain}
The sector sizes of the prediction chain with 3 consecutive steps
of MLPs are visualized in Figure~\ref{fig:predictionchain}. In the
first step the combined prediction of the MLP with topological
input distribution and the MLP with TS ids as additional input
nodes reaches a $z$-resolution of $6.6\,\mathrm{cm}$ and a
$\theta$-resolution of $5.1^\circ$. Accordingly, the sector size in
$\theta$ is set to $30.5^\circ$ for the second step, leading to a
total number of 7 overlapping sectors. The $z$-ranges for the
second step are set to $[-15,15]\,\mathrm{cm}$, in accordance with
the requirement that at most 1\,\% of the events from the
interaction region are predicted outside of the new ranges.

In the second step the combined prediction reaches a
$z$-resolution of $3.2\,\mathrm{cm}$ and a $\theta$-resolution of
$2.8^\circ$, with 0.7\,\% of the test events predicted in the
wrong $\theta$-sector. The sector size in $\theta$ is set to
$17.0^\circ$ for the third and last step, leading to a total of
13 overlapping sectors. The $z$-ranges for the third step are set
to $[-8,8]\,\mathrm{cm}$.

The combined prediction of the third step finally reaches a
$z$-resolution of $1.7\,\mathrm{cm}$ and a $\theta$-resolution of
$1.7^\circ$, with 1\,\% of the test events predicted in the wrong
$\theta$-sector. The test proves that the required $z$-resolution
can be achieved with our concept, starting from a sector with only
2D information and predicting $z$ and $\theta$.

\subsection{Efficiency Analysis}
A good way to illustrate the capabilities of the sector prediction
methods is to look at their Receiver Operating Characteristic
(ROC), which is the true positive rate or efficiency $\varepsilon$
 vs. the false positive rate $r_\mathrm{FP}$ (background contamination),
where ``positives'' are events predicted within a certain range
$[-z_\text{cut},z_\text{cut}]$ on the $z$-axis. True positives
would be physics events from the interaction region ($z=0\,\mathrm{cm}$)  %interesting physics events 
predicted in these ranges,
while false positives are background events predicted in these
ranges. Since we are working with simulated single tracks rather
than real events, we define all tracks within
$z \in [-1,1]\,\mathrm{cm}$ as potentially interesting events
(see Table~\ref{tab:roc}).
The efficiency $\varepsilon$ and contamination
$r_\mathrm{FP}$ are then defined as

\begin{table}
\centering
\caption{Definitions for the Receiver Operating Characteristic}
\begin{tabular}{c|cc}
 & $z_\text{true} \in [-1,1]\,\mathrm{cm}$ & $z_\text{true} \notin [-1,1]\,\mathrm{cm}$\\
\hline
$z_\text{pred} \in [-z_\text{cut},z_\text{cut}]$ & TP & FP\\
$z_\text{pred} \notin [-z_\text{cut},z_\text{cut}]$ & FN & TN
\end{tabular}

\vspace{\baselineskip}
TP: True Positives, FP: False Positives,\\ FN: False Negatives, TN: True Negatives,\\ $z_\text{true}$: true $z$-vertex provided by the simulation,\\ $z_\text{pred}$: prediction of the neural network
\label{tab:roc}
\end{table}

\begin{figure}[!t]
\centering
\def\svgwidth{\columnwidth}
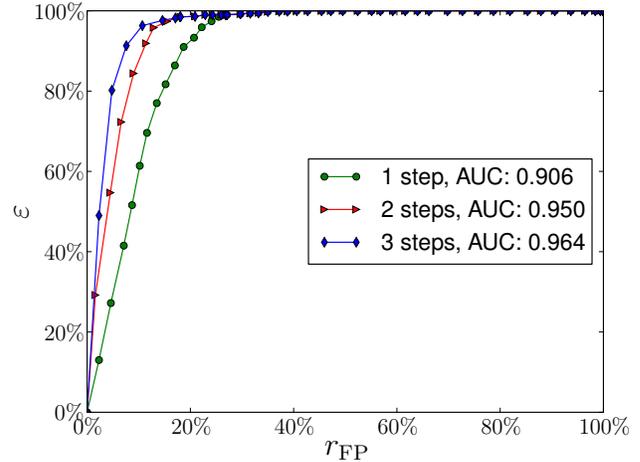
\caption{ROC for $z$ for 3 steps of the prediction chain. The Area Under the Curve (AUC) is a measure for the quality of the predictor, where a perfect predictor would have an AUC of 1.}
\label{roccurve}
\end{figure}

\begin{align}
&\varepsilon = \frac{\text{TP}}{\text{TP}+\text{FN}}; &r_\mathrm{FP} = \frac{\text{FP}}{\text{FP}+\text{TN}}
\end{align}
The parameter $z_\text{cut}$ is varied to generate different
values for the efficiency and the contamination. The
resulting ROC curves for the 3 training steps are shown in
Fig.~\ref{roccurve}. The improvement of the prediction with each
new step is clearly visible. 
With a cut at $z_\text{cut} = 6\,\mathrm{cm}$ a contamination of 17.1\% can be obtained while keeping an efficiency of
98.2\%. Note that the background contamination contains events outside
of the interesting region of $[-1,1]\,\mathrm{cm}$, but within the
cut interval of $[-6,6]\,\mathrm{cm}$, i.e. an MLP with perfect
prediction would still have a false positive rate of 10.2\% at
this cut value. The real background rejection in the experiment will depend
on the topological distribution of the background tracks, 
for which the study has just been started.%which is

\section{Conclusion}
We have presented the concept for a first level $z$-vertex trigger for the Belle~II experiment, using the hit information from the CDC without explicit track reconstruction. The concept is 
based on an ensemble of ``expert'' MLPs, specialized to
sectors in phase space, defined by $\phi, \theta$ and $p_T$, which are small enough to contain single tracks. %The high resolution that can be achieved
Using the $p_T$ and $\phi$ information from the standard CDC 2D trigger, MLPs are used in a preprocessing step to roughly estimate the polar angle $\theta$ and the $z$-vertex for each track. Then the pre-trained  expert MLPs determine the $z$-vertices of the tracks.
Based on our preliminary investigations,
the current state-of-the-art hardware is able to provide
sufficient computing resources to implement the large number of
``expert'' neural networks. In the near future we will explore the
optimal full prediction chain and provide measurements with the
hardware. To this end, new algorithms for the sector finding will
be explored. Especially, the proper combination of different
predictors in each preprocessing step will be optimized.
Then MLPs will be trained for all sectors and the neural
trigger will be tested with full events. The $z$-vertices of
several tracks will be estimated separately and combined to
improve the prediction for the full event. After a successful completion of the studies,
the plan is to install the neural network trigger into the
Belle~II trigger system.

While the presented methods are developed and tested
primarily for the Belle~II experiment, the concept could also be
applied for the development of other track-based trigger
algorithms. 
The proposed method could be useful for the LHC 
experiments in the muon trigger area, estimating e.g. the muon transverse momentum.  
For the future generations of silicon detectors it could also be used
as a secondary vertex trigger. %In the Panda experiment t
The method could even be used for a fast online track reconstruction using, e.g. the straw tube trackers of the Panda experiment at FAIR~\cite{PandaStrawTube}. % input from the straw tube trackers. 
Furthermore, the combination of a
preprocessing step with sectorized local experts is a very 
general divide and conquer approach that can also be
transferred to various other problems.

% or


\appendices

%\section{}

\section*{Acknowledgment}

The simulations have
been carried out on the computing facilities of the Computational
Center for Particle and Astrophysics (C2PAP) within the Excellence Cluster Universe. 
The authors are grateful for the support by F.~Beaujean and J.~Mitrevski through C2PAP.

\ifCLASSOPTIONcaptionsoff
  \newpage
\fi



\bibliographystyle{IEEEtran}
\bibliography{references}
\end{document}